\documentclass{JHEP3}
\usepackage{bm,amsmath,yfonts,epsfig,latexsym}
\usepackage{verbatim, marvosym}
\usepackage[amssymb,cdot]{SIunits}
\newcommand{\qn}{\textswab{q}}
\newcommand{\wn}{\textswab{w}}

\renewcommand{\d}{\partial}

\newcommand{\x}{\bm{x}}

\newcommand{\gYM}{g_{\mathrm{eff}}}
\newcommand{\rsh}{{r_{\mathrm{h}}}}
\def\ofo{ { {}_2 \! F_1 }}
\newcommand{\stru}{\rule[-.2in]{0in}{.2in}}

\def\P{P} 
\def\wf{Z} 

\def\Go{\gamma_0} 
\def\Gr{\gamma_r} 

\title{Holography and hydrodynamics: diffusion on stretched horizons}

\author{Pavel Kovtun\\
Department of Physics, University of Washington,  
Seattle, WA 98195-1560, USA \\
Email: \email{pkovtun@phys.washington.edu}
}
\author{Dam T.~Son and Andrei O.~Starinets\\
Institute for Nuclear Theory, University of Washington,
Seattle, WA 98195-1550, USA\\
Emails: \email{son@phys.washington.edu, starina@phys.washington.edu}
}

\date{June 2003}

\abstract{We show that
 long-time, long-distance fluctuations of plane-symmetric horizons exhibit
universal hydrodynamic behavior. By considering classical fluctuations around
black-brane backgrounds, we find both diffusive and shear modes.
The diffusion constant and the shear viscosity 
are given by simple formulas, 
in terms of metric components. 
For a given metric, the answers can be interpreted 
as corresponding kinetic coefficients in the holographically dual theory.  
For the near-extremal D$p$, M2
and M5 branes, the computed kinetic coefficients coincide with the 
results of independent AdS/CFT calculations.
In all the examples, the ratio of shear viscosity to entropy density is
 equal to $\hbar/(4\pi k_B)$, suggesting a
special meaning of this value.}


\preprint{ UW/PT 03-21 \\ INT-PUB 03-17 \\ hep-th/0309213 }

\begin{document}
\section{Introduction}

The holographic principle
\cite{'tHooft:gx,Susskind:1994vu,Bousso:2002ju} is believed to be
fundamental to the construction of the quantum theory of gravity.  It
asserts that the physics in a region of space, in a theory with
gravity, can be described by a set of degrees of freedom (some ``dual
theory'', which does not contain gravity), associated with a
hypersurface of smaller dimension in that space (the ``holographic
screen'').  The nature of the degrees of freedom on the holographic
screen is not specified by the holographic principle.  In the explicit
example of holography discovered in string theory --- the AdS/CFT
duality
\cite{Maldacena:1997re,Gubser:1998bc,Witten:1998qj,Aharony:1999ti},
the holographic degrees of freedom form a local quantum field theory
(${\cal N}=4$ supersymmetric Yang-Mills theory in four
dimensions), but in general this need not be the case.

In this paper we elucidate the infrared properties of theories whose
gravity duals contain a black brane with a nonzero Hawking
temperature.  We argue that the infrared behavior of these theories is
governed by nothing other than {\em hydrodynamics}.  Namely, by
considering fluctuations around generic black brane solutions we
found modes with dispersion relations of the type $\omega=-iDq^2$,
which are naturally interpreted as diffusion of conserved charges and damped
hydrodynamic shear flow in the holographically dual theory.\footnote{
The gravity dual of the sound wave,
with the dispersion relation $\omega=v_sq-i\gamma q^2$, should also
exist, but is beyond the scope of this paper.}

This result can be anticipated from the holographic principle.
Indeed, the dual theory is at a finite temperature, equal to the
Hawking temperature associated with the gravitational background, 
and is translationally invariant. 
On general grounds, we expect the infrared 
behavior of such a theory to be governed by hydrodynamics. 
The existence of the diffusive modes is consistent with this expectation.

Some hydrodynamic-like properties of event horizons
 have been known for a
long time in the framework of the ``membrane paradigm''
\cite{Thorne:iy,Parikh:1997ma}.  In this approach the stretched
horizon  is interpreted as a fluid with certain
dissipative properties such as electrical conductivity and shear and bulk 
viscosities. Normally, the ``membrane paradigm'' is applied to black holes.
However, for  black holes the relation to hydrodynamics does not go 
beyond the
level of analogy, in particular because their horizons lack 
translational invariance.
The connection to hydrodynamics is much more direct in the case of
black branes. The membrane paradigm suggests the mapping
between the bulk fields and the currents, whereas 
the long wavelength limit has to be taken to derive the hydrodynamic 
equations.

Using the AdS/CFT correspondence, it has been shown  
\cite{Policastro:2001yc,PSShydro,Herzog,PSShydro2,Herzog:2003ke}  that 
the theories living on non-extremal D3, M2 and M5 branes behave
hydrodynamically in the infrared.
In this paper, we will see hydrodynamic behavior
 emerging as a common feature 
of black brane backgrounds. 
Our calculations do not rely on a particular realization of holography 
such as AdS/CFT.
We find explicit formulas expressing the transport
coefficients (diffusion rate, shear viscosity) in terms of the
components of the metric for a wide class of metrics.  

The paper is organized as follows.  In section~\ref{secone} we
consider small fluctuations of black brane backgrounds, and
demonstrate the emergence of Fick's law and diffusion from Maxwell's
equations.  
We derive
an explicit formula for the diffusion constant.  In section~\ref{shear_mp} we
extend the discussion to gravitational perturbations and compute the
shear viscosity.  In section~\ref{applic} the derived formulas are
applied to various supergravity backgrounds.  In the D3, M2, and M5
examples, the answers coincide with the kinetic coefficients in the
dual theories, computed previously by AdS/CFT methods.  In
section~\ref{ads} the kinetic coefficients in the theories dual to
supergravity on D$p$ brane backgrounds are computed using the AdS/CFT
recipe.  The results coincide with those obtained in
section~\ref{applic}.  Section~\ref{sec:concl} discusses some unsolved
problems and possible directions for future work.


\section{Diffusive and shear modes from horizon fluctuations}
\label{secone}
\subsection{Backgrounds}
We start by considering linearized 
perturbations of a black-brane 
background.
We take the $D$-dimensional background metric $G_{MN}$ to be of the form %
\footnote{
     Black brane metrics are solutions to the low-energy string- or M-theory 
     (ten or eleven-dimensional supergravity)
     equations of motion. Full solutions also include other tensor fields 
     such as the Ramond-Ramond field and/or the dilaton.
     However, to the linear order in gravitational perturbations, 
     the presence of 
     these extra fields does not affect subsequent discussion.}
%
\begin{equation}\label{full membrane}
  ds^2 = G_{00}(r)\,dt^2 + G_{rr}(r)\,dr^2 + 
         G_{xx}(r) \sum_{i=1}^{p} (dx^i)^2 
         + \, \wf(r) \, K_{mn}(y) \, dy^m dy^n\,,
\end{equation}
where the components $G_{00}(r)$, $G_{rr}(r)$, $G_{xx}(r)$,  
and the ``warping factor'' $\wf(r)$
depend only on the radial coordinate $r$.\footnote{
       Our choice of metric excludes rotating branes. In other words, 
       we do not consider nonzero chemical potentials.}
We assume that the metric has a plane-symmetric horizon at $r=r_0$,
that extends 
in $p$ infinite spatial dimensions  parametrized by the coordinates $x^i$.
In the vicinity of the horizon $G_{00}$
vanishes, $G_{rr}$ diverges, and $G_{xx}$,
$Z$ stay finite.
The coordinates $y^m$ parametrize some 
$d$-dimensional compact space.

We will be interested in small fluctuations 
of the black brane background,
$G_{MN} + \tilde G_{MN}$,
where $\tilde G_{MN}$ is a
small perturbation, and indices $M$, $N$
run from $0$ to $(1+p+d)$. 
As usual, such perturbations can be
divided up into ``scalar'', ``vector'', and
``tensor'' parts: $\tilde G_{mn}$, $\tilde G_{\mu n}$,
and $\tilde G_{\mu\nu}$, correspondingly.
Here indices $\mu$, $\nu$ run from $0$ to
$p+1$ (labeling the ``non-compact'' 
coordinates $t$, $x^i$ and $r$), 
and indices $m$, $n$ run from $1$
to $d$ (labeling the ``compact'' 
coordinates $y^m$).
For our purposes we will be interested in
vector and tensor perturbations only;
this is motivated by the fact that effective
hydrodynamics in field theories is 
constructed in terms of conserved currents 
and energy-momentum tensor of the theory.

As in the Kaluza-Klein mechanism,
the problem of analyzing $y$-independent 
tensor/vector perturbations reduces to
the problem of gravitational/vector fields
propagating on a lower-dimensional background.
We write the metric of the dimensionally reduced 
background as
\begin{equation}\label{membrane_metric}
  ds^2 = g_{00}(r)\,dt^2 + g_{rr}(r)\,dr^2 + 
         g_{xx}(r) \sum_{i=1}^{p} (dx^i)^2 \ .
\end{equation}
Its components $g_{\mu\nu}$ are just
$G_{\mu\nu}$ of (\ref{full membrane}),
multiplied by $\wf(r)^{d/p}$.
(Details of dimensional reduction are sketched in
 Appendix~\ref{app:dimensional reduction}.)

We will restrict ourselves to the linearized
equations of motion for $\tilde G_{\mu\nu}$ and
$\tilde G_{\mu n}$ on the background
(\ref{membrane_metric}).
The dynamics of the vector perturbations 
is governed by Maxwell's action
\begin{equation}
\label{eq:gauge action}
  S_{\rm gauge} \sim  \int d^{\, p+2}x\, 
                      \sqrt{-g} \,
                      \left(\frac{1}{\gYM^2} \, 
                      F^{\mu\nu} F_{\mu\nu}\right) \ ,
\end{equation}
where $F_{\mu\nu} = \d_\mu A_\nu - \d_\nu A_\mu$, with
the gauge field $A_\mu$ being proportional to  
$\tilde G_{\mu n}$.
In (\ref{eq:gauge action}), $\gYM$ is an
effective coupling, whose overall normalization 
will not be important. 
If $\wf(r)$ is not constant
(the compact space does not factorize),
then $\gYM^2$ is $r$-dependent, and is
proportional to $\wf(r)^{-(p+d)/p}$.

The dimensionally reduced 
metric (\ref{membrane_metric})  has an
event horizon at $r=r_0$, in the vicinity 
of which $g_{00}$
vanishes and $g_{rr}$ diverges,
\begin{subequations}
\begin{eqnarray}
  g_{00} & = &  -(r-r_0)\,\Go + O((r-r_0)^2)\,, \\
  g_{rr} & = & \frac{\Gr}{r-r_0} + O(1)\,.
\end{eqnarray}
\end{subequations}
Here $\Go$ and $\Gr$ are some positive constants.  
We assume  that $g_{xx}(r)$ is a slowly varying function of $r$ 
which does not 
vanish or diverge at the horizon, and is of the same order as 
$g_{00}$ when  $r-r_0\gtrsim r_0$.
%
The Hawking temperature associated with the 
dimensionally reduced background
(\ref{membrane_metric}) is
equal to the Hawking temperature of 
the full metric (\ref{full membrane}),
and is given by
\begin{equation}
\label{eq:membrane T}
    T=\frac{1}{4\pi} \sqrt{\frac{\Go}{\Gr}} \,.
\end{equation}

We start by analyzing vector perturbations.
Tensor perturbations $\tilde G_{\mu\nu}$
will be considered in the next section.

\subsection{Current}

We shall illustrate the appearance of hydrodynamics on a simple
problem of an Abelian gauge field on the 
black-brane background (\ref{membrane_metric}).  
%
Maxwell's equations, which follow from
the action (\ref{eq:gauge action}), read
\begin{equation}
   \d_\mu \left( \frac{1}{\gYM^2}\, \sqrt{-g} \, F^{\mu\nu} \right) = 0 \;.
\end{equation}
To simplify the expressions, we will 
take $\gYM$ to be constant; thus Maxwell's 
equations become 
$\d_\mu \left( \sqrt{-g} \, F^{\mu\nu} \right) = 0$.
The results
for the $r$-dependent $\gYM$ can be easily 
restored by replacing 
$\sqrt{-g} \to \sqrt{-g}/\gYM^2$
in the final answers.

We will be considering fluctuations of 
the gauge field, which behave as
$A_\mu \propto e^{-i\omega t + i{\bm q} \cdot {\bm x}}$.
Equations of motion for the field $A_{\mu}$,
together with appropriate boundary 
conditions give the dispersion law 
$\omega = \omega(q)$.
In the limit ${\bm q\to 0}$,
the dispersion law of the form 
$\omega(q) = -i D {\bm q}^2$,
with $D$ constant, is a sign of 
diffusion.
We will be looking for the diffusive 
dispersion law on the black brane 
background (\ref{membrane_metric}).


In principle, one can derive the dispersion law 
by directly solving Maxwell's equations. 
Here we will follow a somewhat different path.  
A conserved ``current'' $j^\mu$ can be defined
directly in terms of the field strength 
$F_{\mu\nu}$.
One can show that 
if the gauge field satisfies the
equations of motion with relevant boundary conditions, then 
at long distances Fick's law $j^i=-D \, \d_i j^0$ is valid.
Then current conservation $\d_\mu j^\mu = 0$ implies that
the charge density $j^0$ satisfies the diffusion equation 
$\d_t \, j^0=D \, \d_i \d_i \, j^0$.  
%
%

To define the current we need to introduce the {\em
stretched horizon}~\cite{Thorne:iy}.
In our case, the stretched horizon is just a flat spacelike surface located
at a constant $r=\rsh$ slightly larger than $r_0$:
\begin{equation}
  \rsh > r_0\,,  \qquad  \rsh-r_0 \ll r_0 \,.
\end{equation}
The outward normal to that surface is a spacelike vector $n_\mu$,
which has only one nonzero component $n_r = g^{1/2}_{rr}(\rsh)$, so
that $n^2=g^{rr}n_rn_r=1$.
The current which is associated with the stretched horizon is defined 
as in the original membrane paradigm approach
\cite{Thorne:iy,Parikh:1997ma}:
\begin{equation}
  j^\mu = \left. n_\nu F^{\mu\nu}\right|_\rsh \,.
\label{eq:current_definition}
\end{equation}
The antisymmetry of $F^{\mu\nu}$ implies $n_\mu j^\mu=0$.
That is, the current is parallel to the horizon.
Further, Maxwell's equations, and the fact that
all metric components $g_{\mu\nu}$ depend
on $r$ only, imply that $j^\mu$ is conserved, 
$\d_\mu j^\mu=0$, for any choice of $\rsh$.
The components of the current in 
the vicinity of the horizon are
\begin{subequations}
\begin{eqnarray}
  j^0 &=& F^{0r} n_r =  -\frac{F_{0r}} {\Go \Gr^{1/2} (\rsh-r_0)^{1/2}}\,,
  \label{j0F0r}\\
  j^i &=& F^{ir} n_r = \frac{(\rsh-r_0)^{1/2}}{g_{xx}\Gr^{1/2}}F_{ir}\,.
  \label{jiFir}
\end{eqnarray}
\end{subequations}
The diffusion equation can be derived if we can prove Fick's law.  We
shall do that in two steps:
\begin{itemize}
\item From the incoming-wave boundary conditions we infer a relation
between electric and magnetic fields near the boundary:
\begin{equation}\label{incoming-bc}
  F_{ir} =
    -\sqrt\frac{\Gr}{\Go} \frac{F_{0i}}{r-r_0}\,,
  \qquad r-r_0\ll r_0 \,.
\end{equation}
This relation is analogous to the relation 
$\mathbf{B} = -\mathbf{n} \times \mathbf{E}$
for plane waves on a non-reflecting surface 
in classical electrodynamics.
\item We then show that if $A_\mu$ varies slowly in space and time, then
$F_{0i}$ is proportional to $\d_i F_{0r}$ 
(analogously to $\mathbf{E}=-{\bm\nabla}\varphi$ 
in electrostatics), for appropriate choices of
$\rsh$. Then the relations (\ref{j0F0r}), 
(\ref{jiFir}) imply $j^i=-D \, \d_i j^0$, 
and give the expression for $D$.
\end{itemize}
We now present details of the derivation.

\subsection{Boundary conditions}

Because of translation invariance of the metric (\ref{membrane_metric}), 
we can restrict ourselves to field configurations which are
plane waves with respect to the spatial coordinates $x^i$.  
Without loosing generality, we choose the wavevector
to lie along the $x\equiv x^1$ axis, so
\begin{equation}\label{ansatz}
   A_\mu = A_\mu(t,r) e^{iqx}.
\end{equation}
In the hydrodynamics limit $q$ is arbitrarily small.  In particular we will
assume it is much smaller than the Hawking temperature: $q\ll T$.

Fick's law $j^i=-D\,\d_i j^0$ violates time reversal.  Its origin thus
must be rooted in the irreversible nature of the horizon.  To
incorporate time irreversibility we impose the incoming-wave boundary
conditions on the horizon: waves can be absorbed by the horizon but
cannot be emitted from there. 

To derive the relation between electric and magnetic fields, 
we first write down the relevant Maxwell's equations and
Bianchi identities, taking into account the form (\ref{ansatz})
\begin{subequations}
\begin{eqnarray}
  & & g^{00} \d_t F_{0r} - g^{xx} \,\d_x F_{rx} = 0\,,\label{maxwell11}\\
  & & \sqrt{-g} \, g^{00} g^{xx} \d_t F_{0x} 
      + \d_r (\sqrt{-g} \, g^{rr} g^{xx} 
        F_{rx}) = 0\,, \label{maxwell2}\\
  & & \d_r(\sqrt{-g}\,g^{rr}g^{00}F_{0r}) +
      \sqrt{-g}\,g^{xx}g^{00}\,\d_x F_{0x} = 0\,,\label{maxwell3}\\
  & & \d_t F_{rx} + \d_x F_{0r} - \d_r F_{0x} = 0\,. \label{bianchi3}
\end{eqnarray}
\end{subequations}
 Using Eq.~(\ref{bianchi3}) one can write
down field equations containing only the electric fields $F_{0r}$ and
$F_{0x}$,
\begin{subequations}
\begin{eqnarray}
  & & \d_t^2 F_{0r} - g_{00}\,g^{xx} q^2 F_{0r}  
     - (iq) g_{00} \, g^{xx} \d_r F_{0x} = 0\,, \label{F0r-eq}\\
  & & \d_t^2 F_{0x} + g_{00}\,\d_r(g^{rr} \d_r F_{0x})
     - g_{00}\,\d_r (g^{rr}\d_x F_{0r}) = 0\,. \label{F0z-eq}
\end{eqnarray}
\end{subequations}
In the near-horizon region $r-r_0\ll r_0$, these equations simplify
considerably.  Let us assume that fields vary over a typical time
scale $\Gamma^{-1}$, so that $\d_t^2\sim\Gamma^2$.  At small $(r-r_0)$ 
Eq.~(\ref{F0r-eq}) implies
\begin{equation}\label{F0r-negl}
  F_{0r} \sim \frac{r-r_0}{r_0}\frac q{\Gamma^2} \d_r F_{0x}\,.
\end{equation}
Hence at sufficiently small $r-r_0$, namely, when
\begin{equation}\label{r-r0Gammaq}
  \frac{r-r_0}{r_0} \ll \frac{\Gamma^2}{q^2}\,,
\end{equation} 
the third term in the left hand side of Eq.~(\ref{F0z-eq}) is small
compared to the second term.  This equation gives a ``wave equation'' for
$F_{0x}$ alone,
\begin{equation}
  \d_t^2 F_{0x} - \frac{\Go}{\Gr} (r-r_0) \, \d_r
    [ (r-r_0) \d_r F_{0x}] = 0\,.
\end{equation}
The general solution to that equation is
\begin{equation}
  F_{0x}(t,r) = f_1 \left[t+\sqrt{\frac{\Gr}{\Go}}\ln(r-r_0)\right] +
    f_2 \left[t-\sqrt{\frac{\Gr}{\Go}}\ln(r-r_0)\right],
\end{equation}
where $f_1$ and $f_2$ are arbitrary functions.  The incoming-wave
boundary condition picks up the $f_1$ term.  This means
\begin{equation}
  \d_t F_{0x} = \sqrt{\frac{\Go}{\Gr}}(r-r_0) \d_r F_{0x}\,.
\end{equation}
From Eq.~(\ref{bianchi3}) and Eq.~(\ref{F0r-negl}) one finds that
\begin{equation}
   F_{rx} - \sqrt{\frac{\Gr}{\Go}} \frac{F_{0x}}{r-r_0}
\end{equation}
is independent of $t$.  As we expect the solution
to decay as $t\to\infty$, this expression is zero.  Thus we find
Eq.~(\ref{incoming-bc}).

\subsection{Quick derivation of Fick's law}

From now on we will work in the radial gauge $A_r=0$.  We
shall show that the parallel (to the horizon) 
electric field is dominated by the gradient 
of the scalar potential, $F_{0x} \approx -\d_x A_0$, 
and that the normal (to the horizon) electric field
is proportional to the scalar potential itself,
$A_0\sim F_{0r}$.
Hence one finds $F_{0x}\sim \d_x F_{0r}$, 
and therefore $j^x\sim \d_x j^0$.

When $q=0$, $A_0$ satisfies Poisson equation (see
Eq.~(\ref{maxwell3}))
\begin{equation} \label{poisson}
  \d_r \left(\sqrt{-g} \, g^{rr} g^{00} \, \d_r A_0
  \right) =0\,.
\end{equation}
With the boundary condition $A_0(r)=0$ at $r\to\infty$, it
can be solved to yield
\begin{equation} \label{A_0_solution}
  A_0(r) = C_0 \int\limits_r^\infty\!dr'\, 
     \frac{g_{00}(r') \, g_{rr}(r')}{\sqrt{-g(r')}}\,.
\end{equation}
Now let $q$ be nonzero but small.  Our first assumption, (to be
checked later), is that unless $r$ is exponentially
close to $r_0$ one can expand $A_0$ in a series over $q^2/T^2\ll1$, and the
leading term has the same $r$-dependence as in
Eq.~(\ref{A_0_solution}), with $C_0$ depending on coordinates and
time, i.e.,
\begin{equation} \label{A_0_solution1}
\begin{split}
  A_0 &= A_0^{(0)} + A_0^{(1)} + \cdots, \qquad 
  A_0^{(1)} =  O (q^2/T^2)\,,\\
  A_0^{(0)}(t,x,r) &= C_0(t)e^{iqx} \int\limits_r^\infty\!dr'\, 
     \frac{g_{00}(r') \, g_{rr}(r')}{\sqrt{-g(r')}} \,,
\end{split}
\end{equation}
When $r\approx r_0$ (but not exponentially close to $r_0$) the ratio
of $F_{0r}$ and $A_0$ is a constant which can be read off from the
metric,
\begin{equation}
  \left. \frac{A_0}{F_{0r}}\right|_{r\approx r_0} = 
  \frac{\sqrt{-g(r_0)}}{g_{00}(r_0) \, g_{rr}(r_0)}
        \int\limits_{r_0}^\infty\!dr\, 
     \frac{g_{00}(r) g_{rr}(r)}{\sqrt{-g(r)}} \,.
\end{equation}
Note that this ratio is finite since $g_{00}\,g_{rr}$ is finite as $r\to r_0$.
Our second assumption is that 
\begin{equation} \label{assumption}
  |\d_t A_x| \ll |\d_x A_0|\,.
\end{equation}
Now Fick's law arises naturally:
\begin{equation}\renewcommand {\arraystretch}{2}
\label{jidiA0}
\begin{array}{rl}
  {\displaystyle j^x} &   
     {\displaystyle =-\frac{F_{0x}}{g_{xx} \sqrt{\Go(r-r_0)}}} \\  
  & {\displaystyle = \frac{\d_x A_0 }{g_{xx} \sqrt{\Go(r-r_0)}}}\\
  & {\displaystyle = \left(\frac{A_0}{F_{0r}}\right) 
     \frac{\d_x F_{0r} }{g_{xx} \sqrt{\Go(r-r_0)}}}\\
  & {\displaystyle = - D\d_x j^0 \rule{0cm}{0.5cm}}\,,
\end{array}
\end{equation}
where
\begin{equation}\label{D with e constant}
  D = 
  \frac{\sqrt{-g(r_0)}}{g_{xx}(r_0)
        \sqrt{-g_{00}(r_0) \, g_{rr}(r_0)}}
        \int\limits_{r_0}^\infty\!dr\, 
     \frac{-g_{00}(r) \, g_{rr}(r)}{\sqrt{-g(r)}}\,.
\end{equation}
For the gauge field with $r$-dependent 
coupling $\gYM(r)$, the diffusion 
constant is
\begin{equation}\label{D}
  D = 
  \frac{\sqrt{-g(r_0)}}{g_{xx}(r_0) \, \gYM^2(r_0)
        \sqrt{-g_{00}(r_0) \, g_{rr}(r_0)}}
        \int\limits_{r_0}^\infty\!dr\, 
     \frac{-g_{00}(r) \, g_{rr}(r) \, \gYM^2(r)}{\sqrt{-g(r)}}\,.
\end{equation}
The natural estimate for the diffusion constant is $D\sim T^{-1}$.
From Fick's law and continuity it follows that $j^0$ obeys the
diffusion equation, and $C_0(t)\propto e^{-\Gamma t}$, $\Gamma=Dq^2$.  Notice
that (\ref{r-r0Gammaq}) requires that the stretched horizon has to be
sufficiently close to the horizon,
\begin{equation}\label{r-r0qT}
  \frac{\rsh-r_0}{r_0} \ll \frac{q^2}{T^2}\,,
\end{equation}
which is a stronger condition than $\rsh-r_0\ll r_0$.

\subsection{Checking the assumptions}
\label{shorizon}
The short derivation of the diffusion law relies on two assumptions,
Eqs.~(\ref{A_0_solution1}) and (\ref{assumption}).  We now verify that
these assumptions are self-consistent.  Namely, we assume
Eq.~(\ref{A_0_solution1}) and then show that i) (\ref{assumption}) is
valid and ii) (\ref{A_0_solution1}) satisfies field equations.

From Eq.~(\ref{maxwell11})
\begin{equation}
  - \Gamma g^{00} A_0' + i q g^{xx} A_x' =0
\label{a0axnet}
\end{equation}
(where prime denotes derivative with respect to $r$) one can find
$A_x$ from $A_0$.  Combining that with Eq.~(\ref{A_0_solution1}) and
taking into account the boundary condition $A_x|_{r=\infty}=0$, one
finds
\begin{equation}\label{A0x}
\begin{split}
  A_x &= A^{(0)}_x + A^{(1)}_x + \cdots\,,\\
  A^{(0)}_x &= -\frac{i\Gamma}q C_0 e^{-\Gamma t+iqx} 
  \int\limits_r^\infty\! dr'\, 
  \frac{g_{xx}(r') g_{rr}(r')}{\sqrt{-g(r')}}\,.
\end{split}
\end{equation}
For $r-r_0\sim r_0$ we have $A_x/A_0\sim \Gamma/q$.  This comes from
comparing Eqs.~(\ref{A0x}) and (\ref{A_0_solution1}), taking into
account that $g_{00}\sim g_{xx}$ for these values of $r$.  However in
the limit $r\to r_0$ the integral diverges logarithmically since
$g_{rr}\sim (r-r_0)^{-1}$.  For very small $r-r_0$ we then have
\begin{equation}
  A_x \sim A_0 \frac\Gamma q \ln \frac {r_0}{r-r_0}\,.
\end{equation}
Let us look at the condition~(\ref{assumption}).  We have
\begin{equation}
  \frac{|\d_t A_x|}{|\d_x A_0|} \sim \frac{\Gamma^2}{q^2} 
  \ln \frac{r_0}{r-r_0} \,.
\end{equation}
Since $\Gamma^2/q^2 \sim q^2/T^2\ll 1$, the ratio is much smaller than
1 unless $r-r_0$ is exponentially small so that the logarithm is
comparable to $T^2/q^2$.  Therefore (\ref{assumption}) holds if we
choose the location of the stretched horizon $\rsh$ so that
\begin{equation}\label{r-r0range}
  \ln \frac{r_0}{\rsh-r_0} \ll 
  \frac{T^2}{q^2}\,.
\end{equation}
This means $r_h-r_0$ cannot be too small.  Still this condition can be
satisfied simultaneously with (\ref{r-r0qT}).

Now let us check that $A_0$ can be expanded as in Eq.~(\ref{A_0_solution1}).
Expanding Eq.~(\ref{maxwell3}) in series over $q^2$ we find
\begin{equation}
  \d_r\left(\sqrt{-g} g^{rr} g^{00}  \d_r A_0^{(0)}\right) 
  + \d_x\left[\sqrt{-g} g^{xx} g^{00} 
  \left(\d_x A_0^{(0)} - \d_t A_x^{(0)}\right)
    \right] =0 \,.
\end{equation}
Concentrating on $r$ close to $r_0$, and only on the orders of
magnitude of the terms, we find
\begin{equation}
  \d_r^2 A_0^{(1)} \sim \frac{\Gr g^{xx}}{r-r_0} \left(
    q^2 A_0^{(0)} + \Gamma q A_x^{(1)} \right)
  \sim \frac{\Gr g^{xx}}{r-r_0} \left( q^2 + \Gamma^2
  \ln \frac{r_0}{r-r_0} \right)
  A_0^{(0)}\,,
\end{equation}
from which it follows that
\begin{equation}
\begin{split}
  \d_r A_0^{(1)} &\sim \Gr g^{xx} \left( q^2\ln\frac{r_0}{r-r_0}
  + \Gamma^2 \ln^2\frac{r_0}{r-r_0}\right) A^{(0)}_0\\
  &\sim
  \frac1{r_0}\left(\frac{q^2}{T^2}\ln\frac{r_0}{r-r_0} +
    \frac{q^4}{T^4}\ln^2\frac{r_0}{r-r_0} \right) A_0^{(0)}\,,
\end{split}
\end{equation}
where we have used the estimate $g_{xx}\sim \Go r_0$ and $\Gamma\sim
q^2/T$.  We see that $A_0^{(1)}$ is always smaller than $A_0^{(0)}$ for small
$q^2/T^2$.  However, using Eq.~(\ref{a0axnet}) one finds
\begin{equation}
  A_x^{(1)} \sim \frac\Gamma q \left( \frac{q^2}{T^2}\ln^2
   \frac{r_0}{r-r_0} + \frac{q^4}{T^4}\ln^3 \frac{r_0}{r-r_0} \right)
   A_0^{(0)}\,,
\end{equation}
which is smaller than $A_x^{(0)}$ only if the condition (\ref{r-r0range}) 
is satisfied.

Thus, we have established that the assumptions underlying our
derivation of Fick's law are valid, assuming that the location of the
stretched horizon is chosen to satisfy (\ref{r-r0qT}) and
(\ref{r-r0range}).

\section{Shear viscosity}
\label{shear_mp}

So far we have found that small 
fluctuations of the stretched horizon
have properties, which can be viewed 
as corresponding to diffusion of a 
conserved charge in simple fluids.
We now turn to the next simplest 
hydrodynamic mode -- the shear mode.


In principle, it should be possible to define the energy-momentum
tensor $T^{\mu\nu}$ living on the stretched horizon in a manner similar to
Eq.~(\ref{eq:current_definition}) and show that $T^{\mu\nu}$ is
conserved and, with a suitable choice for the location of the stretched
horizon, satisfies the constitutive equations.  We shall
follow a simpler route. We will show that
the corresponding fluctuations of the 
metric obey a diffusive dispersion law,
$\omega = -i {\cal D} q^2$, and identify
${\cal D}$ with $\eta/(\epsilon+\P)$ in the
dual theory.
Here $\eta$ is the shear viscosity, 
$\epsilon$ and $\P$ are the equilibrium
energy density and pressure
(see Appendix~\ref{app:hydro review}).  We will call ${\cal D}$ the
shear mode diffusion constant.

We write fluctuations of the $(p+2)$-dimensional background
(\ref{membrane_metric}) as 
$g_{\mu\nu}+ h_{\mu\nu}$
and consider only those perturbations which depend
on $t$, $r$, $x=x^1$, with only two non-vanishing components
of $h_{\mu\nu}$
\begin{equation}\label{grav_ansatz}
   h_{ty} =  h_{ty}(t,x,r)\,, \qquad
   h_{xy} =  h_{xy}(t,x,r)\,,
\end{equation}
where $x\equiv x^1$, $y\equiv x^2$ (assuming the black brane
is extended along at least two spatial dimensions).
This is motivated by the fact that 
hydrodynamic shear mode describes 
decay of the fluctuations in transverse
momentum density of the fluid.
We shall call perturbations of the type (\ref{grav_ansatz})
gravitational shear perturbations.  

The linearized field equations for $h_{ty}$ and
$h_{xy}$ decouple from all other modes and 
so they can be treated
separately.%
\footnote{
	By a choice of gauge one can set $h_{ry}$ to zero.}
The easiest way to find the field equations for the gravitational
shear perturbations is to notice that since $h_{ty}$ and $h_{xy}$ do not
depend on $y$,
they can be viewed as zero modes 
of the Kaluza-Klein reduction on
a circle along the $y$ direction.
%
%
From the standard
formulas of Kaluza-Klein compactification one finds that the
fields
$A_0 = (g_{xx})^{-1} h_{ty}$,
$A_x = (g_{xx})^{-1} h_{xy}$
satisfy Maxwell's equations on a 
$(p+1)$-dimensional background,
whose metric components are
\begin{equation}
\label{eq:second-reduced metric}
   \bar g_{\alpha\beta} = g_{\alpha\beta} \left(g_{xx}\right)^\frac{1}{p-1}
\end{equation}
The indices $\alpha$, $\beta$ run from $0$ to $p$.
Indeed, the
gravitational action contains the following piece
\begin{equation}\label{KK}
   \sqrt{-g}\, {\cal R}_{(g)} \rightarrow -\frac14 \,
   \sqrt{-\bar g} \,  
   F_{\alpha\beta} F_{\gamma\delta} \,
   \bar{g}^{\alpha\gamma} \bar{g}^{\beta\delta}
   \left(g_{xx}\right)^\frac{p}{p-1} + \dots\,,
\end{equation}
where 
$F_{\alpha\beta} = \d_\alpha A_\beta - \d_\beta A_\alpha$,
and the only non-zero components are
$F_{tx}$, $F_{tr}$, and $F_{xr}$.

Thus the problem is reduced to that of Abelian vector field 
considered in the previous section,%
\footnote{
	When solving for gravitational shear 
	perturbations in $10$ or $11$-dimensional
	supergravity, one can consistently set
	fluctuations of all other fields to 
	zero.}
with the identification 
$1/\gYM^2 = \left(g_{xx}\right)^\frac{p}{p-1}$. 
The shear mode damping constant ${\cal D}$
is given by the result (\ref{D}), 
evaluated with $\bar g_{\alpha\beta}$.
Using the relation (\ref{eq:second-reduced metric})
between $g_{\mu\nu}$ and $\bar g _{\mu\nu}$,
one finds
\begin{eqnarray}
   {\cal D} &=& \frac{\sqrt{-g(r_0)}}{\sqrt{-g_{00}(r_0) \, g_{rr}(r_0)}}
                \int\limits_{r_0}^\infty\!dr\,
                \frac{-g_{00}(r) \, g_{rr}(r)}{g_{xx}(r) \sqrt{-g(r)}} 
                \nonumber \\
            &=& \frac{\sqrt{-G(r_0)}}{\sqrt{-G_{00}(r_0) \, G_{rr}(r_0)}}
                \int\limits_{r_0}^\infty\!dr\,
                \frac{-G_{00}(r) \, G_{rr}(r)}{G_{xx}(r) \sqrt{-G(r)}} \ .
\label{smdcd}
\end{eqnarray}
In writing the last expression, the
relation between components of the 
black brane metric $G_{\mu\nu}$, and its
dimensionally reduced version $g_{\mu\nu}$
was used, $g_{\mu\nu} = G_{\mu\nu} \wf^{d/p}$.

Using thermodynamic relation $\epsilon+\P=Ts$,
one finds shear viscosity
\begin{equation}\label{eta}
  \eta = s\,T \frac{\sqrt{-G(r_0)}}{\sqrt{-G_{00}(r_0) \, G_{rr}(r_0)}}
  \int\limits_{r_0}^\infty\!dr\,
  \frac{-G_{00}(r) \, G_{rr}(r)}{G_{xx}(r) \sqrt{-G(r)}} \,.
\end{equation}
This expression gives shear viscosity in terms 
of components of the black brane metric 
(\ref{full membrane}), its Hawking 
temperature $T$ 
and its entropy $s$ per unit $p$-dimensional
volume.

The formula (\ref{eta}), and the 
corresponding one for the diffusion
constant,
\begin{equation}
  D = 
  \frac{\sqrt{-G(r_0)} \, \wf(r_0)}{G_{xx}(r_0) 
        \sqrt{-G_{00}(r_0) \, G_{rr}(r_0)}}
        \int\limits_{r_0}^\infty\!dr\, 
     \frac{-G_{00}(r) \, G_{rr}(r)}{\sqrt{-G(r)} \, \wf(r)}\,.
\label{main_D}
\end{equation}
are the main results of this paper.

\section{Applications}
\label{applic}

In this section, we shall apply the general formulas
(\ref{eta}) and 
(\ref{main_D}) to different gravitational backgrounds.  
For near-extremal D3, M2
and M5 branes, the results coincide with those found previously in
Refs.~\cite{Policastro:2001yc,PSShydro,Herzog} from the AdS/CFT
correspondence.

\subsection{Near-extremal D3-branes}
The metric of a stack of $N$ non-extremal D3 branes 
in type IIB supergravity is given in the near-horizon region by
\begin{equation}
\label{eq:D3 metric}
  ds^2 = \frac{r^2}{R^2}\left(-f(r)\,dt^2 + dx^2 + dy^2 + dz^2 \right)
  +  \frac{R^2}{r^2f(r)}\,dr^2 + R^2 d\Omega_5^2\,.
\end{equation}
Here $R$ is a constant, which depends 
on the number of D3 branes, 
$R\propto N^{1/4}$, and
$f(r) = 1 - r_0^4/r^4$. 
(This metric is an important example, 
because for type IIB supergravity on 
this background, the holographically
dual theory is known explicitly.)
The Hawking temperature of the 
background metric (\ref{eq:D3 metric}) 
is $T = {r_0}/{\pi R^2}$, and the 
entropy per unit (three-dimensional)
volume is $s = \frac{\pi^2}2N^2T^3$.

Applying Eqs.~(\ref{main_D}), (\ref{eta}), one finds the 
corresponding diffusion constant,%
\begin{equation}
\label{eq:D in SYM}
   D = \frac1{2\pi T}\;,
\end{equation}
and the shear viscosity
\begin{equation}
\label{eq:eta in SYM}
   \eta = \frac{\pi}{8} N^2T^3 \; .
\end{equation}

The holographic dual theory for type IIB 
supergravity on the background (\ref{eq:D3 metric}) 
 is ${\cal N}=4$ supersymmetric Yang-Mills 
theory with gauge group $SU(N)$, in the limit 
of large $N$ and large 't Hooft coupling \cite{Maldacena:1997re}.
This field theory lives in $3+1$ (infinite, 
flat) dimensions, and must be taken at finite 
temperature, equal to the Hawking temperature
of the gravitational background (\ref{eq:D3 metric}).
Long-time, long-distance behavior in 
this theory is governed by conventional 
hydrodynamics,%
\footnote{
     For systematic discussion of hydrodynamic
     fluctuations in supersymmetric theories,
     see \cite{KY}.}
with hydrodynamic variables being just densities
of conserved charges.
This theory has $SU(4)$ global symmetry 
current ($R$-current) which therefore 
must relax diffusively. 
The Fick's law for this current is
${\bm j}_a = -D_{ab} {\bm\nabla}j^0_b$,
where $a,b$ are adjoint $SU(4)$ indices,
which run from 1 to 15.
When thermal equilibrium state respects the
$SU(4)$ symmetry (no chemical potentials
for R-charges), the matrix of diffusion
constants must be an invariant of the group:
$D_{ab} = D_{\!R}\,\delta_{ab}$.
Thus the R-charge diffusion is characterized 
by only one constant $D_{\!R}$, which can be 
computed by using the AdS/CFT approach~\cite{PSShydro}.
The result of the AdS/CFT calculation of $D_{\!R}$
is exactly equal to (\ref{eq:D in SYM}).
Likewise, the result (\ref{eq:eta in SYM})
exactly coincides with the shear viscosity 
of the supersymmetric Yang-Mills plasma, as
computed by AdS/CFT methods 
\cite{Policastro:2001yc, PSShydro}.

In this example, the ratio of shear viscosity to
entropy density is $\eta/s = 1/(4\pi)$.  

\subsection{M2 branes}

The metric of a stack of $N$ non-extremal M2 branes 
in the eleven-dimensional supergravity is given by
\begin{equation}\label{M2}
  ds^2 = H^{-2/3}(-fdt^2+d\x^2)+H^{1/3}(f^{-1}dr^2+r^2d\Omega^2_7)\,,
\end{equation}
where  $H = 1+R^6/r^6$,  $f = 1-r_0^6/r^6$, and
$R$ is a constant, which depends on the number
of M2 branes, $R\propto N^{1/6}$.
The Hawking temperature of the background
metric (\ref{M2}) is
\begin{equation}
   T = \frac{3}{2\pi r_0} \, \frac{1}{H^{1/2}(r_0)}\,.
\end{equation}
Using Eqs.~(\ref{main_D}) and (\ref{eta}), we find
\begin{subequations}
\begin{eqnarray}
  D &=& \displaystyle{\frac{r_0}{8}} \, H^{3/2}(r_0) \; 
\ofo \! \left( 1, \frac43 ; \frac73; - \frac{R^6}{r_0^6}
\right)\,,\stru
  \label{eq:D-M2}\\ 
  \displaystyle{\frac\eta{\epsilon+P}} &=& 
   \displaystyle{\frac{r_0}6} \, H^{1/2}(r_0) = \frac1{4\pi T}\,,
\end{eqnarray}
\end{subequations}
where $\ofo (a,b;c;z)$ is the hypergeometric function. In
the near-extremal limit $r_0\ll R$, 
the expression (\ref{eq:D-M2}) gives%
\footnote{
     One can either use the full metric (\ref{M2})
     in the formula (\ref{main_D}) and then take the 
     near-extremal limit $r_0\ll R$, or directly 
     substitute the near-horizon limit of the metric
     (\ref{M2}) in Eq. (\ref{main_D}), with the same 
     result.}
$D=3/(4\pi T)$.  

The details of the CFT holographically dual  to the near-horizon limit of 
the background (\ref{M2}) 
are not known explicitly.
However, it is known that its symmetry algebra
includes (among other things) translation
symmetry, and a global $SO(8)$ $R$-symmetry.
Thus it makes sense to consider the long-time relaxation
of charge densities of the corresponding conserved 
currents in the dual theory.
The Minkowski AdS/CFT recipe \cite{Son:2002sd}, \cite{Herzog:2002pc}
can be used to calculate correlation functions 
of the energy-momentum tensor and $R$-currents in this 
theory at finite temperature, in the low-frequency limit.
The AdS/CFT calculation \cite{Herzog} shows that the
$R$-current relaxes diffusively, with the diffusion
constant equal to $3/(4\pi T)$, and that the shear
mode damping constant is $1/4\pi T$.
Thus we see again that the general formulas
 (\ref{eta}) and (\ref{main_D}) reproduce previous
AdS/CFT results.

In this example, we also have $\eta/s=(4\pi)^{-1}$, 
independent of the ratio $r_0/R$.

\subsection{M5 branes}

The metric of a stack of $N$ non-extremal M5 branes 
in the eleven-dimensional supergravity is given by
\begin{equation}\label{M5}
  ds^2 = H^{-1/3}(-fdt^2+d\x^2)+H^{2/3}(f^{-1}dr^2+r^2d\Omega^2_4)\,,
\end{equation}
where  $H = 1+ R^3/r^3$,  $f = 1-r_0^3/r^3$, and
$R$ is a constant, which depends on the number
of M5 branes, $R\propto N^{1/3}$.
The Hawking temperature of the background metric (\ref{M5}) is
\begin{equation}\label{hawking_m5}
  T = \frac3{4\pi r_0} \frac1{H^{1/2}(r_0)}\,.
\end{equation}
Applying Eqs.~(\ref{main_D}) and (\ref{eta}), we get 
\begin{subequations}
\begin{eqnarray}
  D &=& \displaystyle{\frac{r_0}{5}}\, H^{3/2}(r_0)\; 
\ofo \! \left( 1, \frac53 ; \frac83 ; - \frac{R^3}{r_0^3}\right)\,,\stru
        \label{eq:D-M5}\\
  \displaystyle{\frac\eta{\epsilon + P}} &=& \displaystyle{\frac{r_0}3} \,
     H^{1/2}(r_0) = \frac1{4\pi T}\,.
\end{eqnarray}
\end{subequations}
  In the near-extremal limit $r_0\ll R$, the expression (\ref{eq:D-M5}) gives
 $D=3/(8\pi T)$.  
Analogously to the previous example of M2 branes,
these values reproduce the AdS/CFT results of
Ref.~\cite{Herzog}.
Again, in this example $\eta/s=(4\pi)^{-1}$.

\subsection{D$p$ branes}
\label{sec:pbranes}

Black $p$-brane metrics are solutions to the low-energy
string theory equations of motion \cite{Horowitz:cd}. 
The metric ($p<7$) in the Einstein frame reads
\begin{equation}\label{es_metric}
  ds_{E}^2 = H^{-\frac{7-p}8}
  \left( -f dt^2 + d x_1^2 + \cdots + d x_p^2\right) 
  + H^{\frac{p+1}8}\left( f^{-1}dr^2 + r^2 d\, \Omega_{8-p}^2\right)\,,
\end{equation}
where 
\begin{equation}
  H = 1 + \frac{R^{7-p}}{r^{7-p}}\,, \qquad 
  f = 1 - \frac{r_0^{7-p}}{r^{7-p}}\,.
\end{equation}
The Ramond-Ramond field strength is given by 
\begin{equation}\label{RR}
  F_{r012\ldots p} = (p-7) \frac{R^{(7-p)/ 2 }
  \sqrt{r_0^{7-p}+R^{7-p}}}{H^2(r)\, r^{8-p}}\,,
\end{equation}
and the dilaton is 
\begin{equation}
e^{\Phi} = H^{(3-p)/4}(r)\,.
\end{equation}
The Hawking temperature is
\begin{equation}\label{hawking}
T = \frac{7-p}{4 \pi r_0} H^{-1/2}(r_0)\,.
\end{equation}
According to the AdS/CFT dictionary, the isometries of the
$(8-p)$-dimensional sphere in the metric correspond to a SO($9-p$)
global R-symmetry of the dual theory.  Fluctuations of the metric with
one index along the brane and another one along the sphere become,
upon dimensional reduction on S$^{8-p}$, gauge fields propagating in
the $(p+2)$-dimensional bulk which couple to the R-symmetry currents
on the $(p+1)$-dimensional flat boundary. 

Dimensionally reducing the full D$p$-brane metric (\ref{es_metric})
we obtain the following metric
\begin{equation}\label{es_metric_dim_red}
  ds_{p+2}^2 = H^{\frac1p} r^{\frac{2(8-p)}p}
  \left( -f dt^2 + d x_1^2 + \cdots + d x_p^2\right) 
  + H^{\frac1p+1}   r^{\frac{2(8-p)}p} f^{-1} dr^2\,.
\end{equation}
The action of the SO($9-p$) gauge field is
\begin{equation}\label{gauge_action}
  S \sim \int\!d^{p+2} x\, \sqrt{-g_{p+2}} \frac1{\gYM^2} F^2_{\mu\nu}\,,
\end{equation}
where constant normalization factors are ignored. The effective gauge 
coupling constant depends on the radial coordinate,
\begin{equation}\label{gauge_eff}
   \gYM^2 (r) = H^{-\frac{p+1}p} r^{-\frac{16}p}\,.
\end{equation}
Applying the formulas (\ref{D}) and (\ref{eta}) we find
\begin{subequations}\label{dcm}
\begin{eqnarray}
  D &=&  
  \displaystyle{\frac{r_0^{8-p}  H^{3/2}(r_0)}{2 R^{7-p}}}\,
  \ofo \left( 1, - \frac2{7-p}; \frac{5- p}{7-p}; 
  - \frac{r_0^{7-p}}{R^{7-p}}\right)\,,\stru\\
  \displaystyle{\frac\eta{\epsilon+P}} &=&
  \frac{r_0}{7-p}  H^{1/2}(r_0) = \frac1{4\pi T}\,.
\end{eqnarray}
\end{subequations}
We find again that $\eta/s=(4\pi)^{-1}$, and in the near-extremal
regime $r_0\ll R$ the R-charge diffusion constant is $D=(7-p)/(8\pi
T)$.  
We shall reproduce these results in
Section~\ref{ads} by using the AdS/CFT prescription.

\subsection{A universal lower bound on $\eta/s$?}

In all examples considered so far, there is a remarkable regularity:
the ratio of shear viscosity to entropy density is always equal to
$(4\pi)^{-1}$.  This ratio is independent of the parameters of the
metric and even of the dimensionality of space-time.  This fact
becomes more interesting if we notice that for any spacetime dimension
the ratio $\eta/s$ has the dimension of the Planck constant.  In the
SI system of units the ratio $\eta/s$ in all cases considered is
\begin{equation}\label{etaovers}
  \frac\eta s = \frac{\hbar}{4\pi k_B} \approx 6.08\times 10^{-13}~
  \kelvin\cdot\second\,.
\end{equation}
One may suspect that the constancy of $\eta/s$ is an inherent property
of classical gravity, and hence Eq.~(\ref{etaovers}) should be valid
for any theory with gravitational dual description.  This includes
${\cal N}=4$ supersymmetric Yang-Mills theory in the regime of large
$N$, large 't Hooft coupling, and its deformations.

We have checked this conclusion on the example of the supergravity
solution recently constructed by Buchel and Liu~\cite{Buchel:2003ah} .
It describes the finite-temperature ${\cal N}=2^*$ SU($N$) gauge
theory at large 't Hooft coupling.  The solution is a non-extremal
generalization of the Pilch-Warner RG flow \cite{Pilch:2000ue}
describing the mass deformation of the ${\cal N}=4$ SYM by a
hypermultiplet term.  In addition to the temperature $T$,  new
parameters, the hypermultiplet masses $m_b$, $m_f$, 
appear in the solution. The
leading corrections to the metric are known in the high-temperature
regime $m_b/T \ll 1$, $m_f/T \ll 1$, 
with $m_b/T=0$, $m_f/T=0$ corresponding to the near-extremal
black three-brane solution.  In Appendix~\ref{app:Buchel-Liu} we show
that the leading corrections to $\eta/s$ vanish identically.

One may suspect that the number $\hbar/(4\pi k_B)$ is somehow special.
We conjecture that it is a lower bound on $\eta/s$.  Since this lower
bound does not contain the speed of light $c$, we suggest that this is
a lower bound for all systems, {\em including non-relativistic ones}.
This means we can check the conjecture on common substances where both
viscosity and entropy density have been measured.

Take, for example, water under normal conditions (298.15 K,
atmospheric pressure).  From the table~\cite{CRC} one finds
$s\approx3.9\times10^6~\joule\cdot\kelvin^{-1}\cdot\metre^{-3}$ and
$\eta\approx0.89~\milli\pascal\cdot\second$.  This means
$\eta/s\approx 2.3\times10^{-10}~\kelvin\cdot\second$, which is by a
factor of 400 larger than~(\ref{etaovers}).  We have checked other
common substances and found that the ratio $\eta/s$ is always larger than
$\hbar/(4\pi k_B)$.  The minimum value we found is for liquid $^4$He
at 1~MPa and 10~K, for which
$\eta\approx3.07~\micro\pascal\cdot\second$ and
$s\approx4.95\times10^5~\joule\cdot\kelvin^{-1}\cdot\metre^{-3}$
\cite{CRC}, and the ratio $\eta/s$ is still larger
than~(\ref{etaovers}) by a factor of 10.

Let us now argue that in weakly coupled theories the ratio $\eta/s$
should be much larger than 1 and thus satisfies the lower bound.
Weakly coupled systems can be described as dilute gases of weakly
interacting quasiparticles.  Finite-temperature $\lambda\phi^4$ theory
with $\lambda\ll1$ and gauge theories with small gauge coupling
$g\ll1$ belong to this class.  Nonrelativistic Bose and Fermi liquids
at temperatures much below quantum degeneracy are also weakly coupled
since they can be described as dilute gases of quasiparticles (phonons
in the case of Bose liquids, and dressed fermionic quasiparticles in
the case of Fermi liquids).

The entropy density $s$ of a weakly coupled system is proportional to
the number density of quasiparticles $n$,
\begin{equation}
  s \sim k_B n \,.
\end{equation}
The shear viscosity is proportional to the product of the energy
density and the mean free time (time between collisions) $\tau$
\begin{equation}
  \eta \sim n\epsilon \tau \,,
\end{equation}
where $\epsilon$ is the average energy per particle (which is of the order
of the temperature $T$).  Therefore
\begin{equation}
  \frac\eta s \sim \frac{\epsilon\tau}{k_B}\,.
\end{equation}
Now, in order for the quasiparticle picture to be valid,
 the width of the quasiparticles must be small compared to their
energies, i.e., one should have
\begin{equation}
  \frac\hbar\tau \ll \epsilon
\end{equation}
which means that
\begin{equation}
  \frac\eta s \gg \frac{\hbar}{k_B}\,.
\end{equation}

Thus we see that (in units $\hbar=k_B=1$) the ratio
$\eta/s$ is large at weak coupling.  Theories whose duals are
described by supergravity are typically strongly coupled, thus
naturally having $\eta/s$ of order 1.  It is still puzzling that this
ratio takes the same value of $(4\pi)^{-1}$ for all such theories.


\section{Transport coefficients from AdS/CFT}
\label{ads}

We have seen that the two simple formulas (\ref{eta}) and (\ref{main_D})
reproduce the known results for the transport coefficients of theories
living on D3, M2 and M5 branes at finite temperature.  For the general
case of D$p$ brane there is no previous calculation to compare our
result with.  Here we show how to find the R-charge diffusion rate
and the viscosity by directly calculating the correlation functions of
the R-current and the stress-energy tensor following the prescription
of Ref.~\cite{Son:2002sd}.  (At zero temperature, such correlators were
considered in Ref.~\cite{Gherghetta:2001iv}.)

The metric of black D$p$-branes is given by Eq.~(\ref{es_metric}).  In
this section we work exclusively in the near-horizon limit $r\ll
R$.  It will be convenient to introduce a new radial variable ($u=r_0/r$
for metrics with even $p$ and $u=r_0^2/r^2$ for the ones with odd $p$),
in terms of which the near-horizon metric for even and odd $p$ respectively
 becomes
\begin{equation}\label{es_metric_u_even}
  ds^2_{E} = \left( \frac{r_0}{uR}\right)^{\frac{(7-p)^2}8} 
  \left( - f dt^2 + d x_1^2 + \cdots + d x_p^2\right) 
  + \left(\frac{uR}{r_0}\right)^{\frac{(7-p)(p+1)}8}
  \frac{r_0^2}{u^2}\left(\frac{d u^2}{u^2 f} 
+ d \Omega^2_{8-p}\right)\,,
\end{equation}
where  $f(u)= 1-u^{7-p}$ (even $p$) and 
\begin{equation}\label{es_metric_u_odd}
ds^2_{E} = \left( {r_0^2\over u R^2}\right)^{{(7-p)^2\over 16}} 
\left( - f dt^2 + d x_1^2 + \cdots + d x_p^2\right) 
+ \left( {u R^2\over r_0^2}\right)^{{(7-p)(p+1)\over 16}}
{r_0^2\over u}\left( {d u^2\over 4 u^2 f}
 + d \Omega^2_{8-p}\right)\,,
\end{equation} 
where  $f(u)= 1-u^{{7-p\over 2}}$ (odd $p$).

The dimensionally reduced metric (\ref{es_metric_dim_red}), 
correspondingly, takes the form 
\begin{equation}\label{es_metric_u_dr_even}
ds^2_{E} = R^{\frac{7-p}p}
\left(\frac{r_0}u\right)^{\frac{9-p}p} 
\left( - f dt^2 + d x_1^2 + \cdots + d x_p^2\right) 
+ R^{6-p+7/p} \left(\frac{r_0}u\right)^{p+\frac9p-6}
 \frac{d u^2}{u^2 f}\,
\end{equation}
for even $p$ and 
\begin{equation}\label{es_metric_u_dr_odd}
ds^2_{E} = R^{{7-p\over p}}
\left( {r_0^2\over u }\right)^{{9-p\over 2 p}} 
\left( - f dt^2 + d x_1^2 + \cdots + d x_p^2\right) 
+ R^{6-p+7/p} \left( {r_0^2\over u}\right)^{{p\over 2} +{9\over 2 p}-3}
 {d u^2\over 4 u^2 f}\,
\end{equation}
for odd $p$, with the same functions $f(u)$ as in 
Eqs.~(\ref{es_metric_u_even}), (\ref{es_metric_u_odd}).

For the backgrounds (\ref{es_metric_u_even}), (\ref{es_metric_u_odd}) 
 we now compute the R-current
retarded correlators in dual gauge theories, as well as the
correlators of the shear-mode components of the stress-energy
tensor. For frequencies and momenta much smaller than the temperature
we find that each of these correlators has a diffusion-type pole with
a specific value of the diffusion constant. The calculations are very
similar to those done for the D3-brane background in
Ref.~\cite{PSShydro} and for M-branes in Ref.~\cite{Herzog}, where
details of the approach can be found.  Our results are summarized in
Table~\ref{tabDD}.

\TABULAR{|c|ccc||c|ccc|}{
\hline
Brane & $s$ & $\displaystyle{\frac\eta{\epsilon+P}}$\rule[-.2in]{0in}{.2in} 
\rule[.14in]{0in}{.14in}
& $D$ & 
Brane & $s$ & $\displaystyle{\frac\eta{\epsilon+P}}$ & $D$\\ 
\hline & & & &&&&\\
D1 & $\sim \displaystyle{\frac{N^2 T^2}{\sqrt{g^2_{YM} N}}}$ & --- 
   & $\displaystyle{\frac3{4 \pi T}}$ &
D5 & $s = \displaystyle{\frac\epsilon T}$
   & $\displaystyle{\frac1{4 \pi T}}$ 
   & $\displaystyle{\frac1{4 \pi T}}$ \\
&  & & &&&& \\
D2 &  $\sim \displaystyle{\frac{N^2 T^{7/3}}{(g^2_{YM} N)^{1/3}}}$ 
   &  $\displaystyle{\frac1{4 \pi T}}$ 
   & $\displaystyle{\frac5{8 \pi T}}$ &
D6 & $\sim \displaystyle{\frac{N^2}{(g^2_{YM}N)^3 T^3}}$ 
   & $\displaystyle{\frac1{4 \pi T}}$ 
   & $\displaystyle{\frac1{8 \pi T}}$\\  
& & & &&&&\\
D3 &  $\sim N^2 T^{3}$ & $\displaystyle{\frac1{4 \pi T}}$ 
   & $\displaystyle{\frac1{2\pi T}}$ & 
M2 & $\sim N^{3/2} T^2$ & $\displaystyle{\frac1{4 \pi T}}$ 
   & $\displaystyle{\frac3{4 \pi T}}$ \\
 & & & &&&& \\
D4 &  $\sim g^2_{YM} N^3 T^{5}$ &  $\displaystyle{\frac1{4 \pi T}}$
   & $\displaystyle{\frac3{8\pi T}}$ &
M5 & $\sim N^{3} T^5$ & $\displaystyle{\frac1{4 \pi T}}$ 
   & $\displaystyle{\frac3{8 \pi T}}$ \\  
& & & &&&&\\ \hline
}
{Entropy 
density \cite{Klebanov:1996un} 
and diffusion constants for black $p$- and M-branes. There is 
no gravitational shear mode for D1 brane. For D5 brane, the entropy
is linear in energy.
\label{tabDD}}

\subsection{R-charge diffusion}
\label{R_charge}

In computing two-point functions of R-currents in the large $N$ limit,
it is sufficient to treat the bulk gauge fields as Abelian ones,
ignoring the self-interactions~\cite{Chalmers:1998xr}.  Accordingly,
to find the R-charge diffusion mode, we consider Maxwell's equations
\begin{equation}\label{maxwell_eq}
\frac1{\sqrt{-g}} \partial_{\nu} \left( \sqrt{-g} g^{\mu\rho}  g^{\nu\sigma}
F_{\rho\sigma}\right) = 0
\end{equation}
on the backgrounds (\ref{es_metric_u_dr_even}), (\ref{es_metric_u_dr_odd}).  
Here to simplify notations we denote $
\sqrt{-g} \equiv \sqrt{-g_{p+2}}/\gYM^2$.  We choose the gauge
$A_u=0$ and restrict ourselves to fields which depend only on the $u$,
$t$ and $x\equiv x^1$,
\begin{equation}
   A_{\mu}(u, t,z) = \int\! \frac{d\omega\, dq}{(2\pi )^2}\, 
   e^{-i\omega t + i q x}
   A_{\mu} (\omega,q,u)\,.
\label{fourier}
\end{equation}
Then for the components 
$A_t$ and $A_x$ one finds the system of equations
\begin{subequations}
\begin{eqnarray}
 & & g^{tt} \omega A_t' - q g^{xx} A_x' =0\,,\label{max1}\\
 & & \d_u \left( \sqrt{-g} g^{tt} g^{uu} A_t'\right) - 
 \sqrt{-g} g^{tt} g^{xx} \left( \omega q A_x + q^2 A_t \right) =0\,,
  \label{max2}\\
  & & \d_u \left( \sqrt{-g} g^{xx} g^{uu} A_t'\right) - 
 \sqrt{-g} g^{tt} g^{xx} \left( \omega q A_t + \omega^2 A_x \right) =0\,,
  \label{max3}
\end{eqnarray}
\end{subequations}
where prime denotes the derivative with respect to $u$.
  The equations for other components of
$A_{\mu}$ decouple, and thus these components can be consistently set
to zero.  Eqs.~(\ref{max1}) and (\ref{max2}) can be combined to give a
closed-form equation for $A_t'$
\begin{equation}\label{max_combined}
\frac d{du} \left[ \frac{\partial_u ( \sqrt{-g} g^{tt} g^{uu} A_t')} 
{\sqrt{-g} g^{tt} g^{xx}}\right] - 
\left(\frac{g^{tt}}{g^{xx}} \omega^2 + q^2\right) A_t' =0\,.
\end{equation}
According to the method developed in Ref.~\cite{PSShydro}, one solves
Eq.~(\ref{max_combined}) for $A_t'$ with the incoming-wave boundary
condition at $u=1$.  This can be done in the hydrodynamic regime
$\omega/T \ll 1$, $q/T \ll 1$.  The obtained $A_t'$ depends on a
single overall normalization constant which can be fixed from
Eq.~(\ref{max2}) and the boundary values $A_t^0 = \lim_{u\rightarrow
0} A_t(u)$, $A_z^0 = \lim_{u\rightarrow 0} A_z(u)$.  One then finds
$A_x'$ from Eq.~(\ref{max1}).  This is sufficient for computing the
on-shell classical action, which is proportional to $A_tA_t' - A_x A_x'$ at
$u=0$.  Taking derivatives of the action with respect to $A_t^0$ and
$A_x^0$ one then obtains the current correlators.

In backgrounds (\ref{es_metric_u_dr_even}), (\ref{es_metric_u_dr_odd}), 
Eq.~(\ref{max_combined}) becomes

\begin{equation}\label{com_odd_1}
{d\over d u} \left[\, (1-u^{\nu})\, u^{\alpha }\, {d\over d u}
 \left( u^{\beta } A_t'\right)\, \right] + \left( {\wn^2\over 1-u^{\nu}}
 - \qn^2\right) A_t' =0\,,
\end{equation}
where for $p$ even
\begin{equation}\label{param_R_even}
 \nu_{even} = 7 - p\,, \;\;\;\; \alpha_{even}
  =  p-2\,, \;\;\;\;
\beta_{even}  = -1\,,
\end{equation}
while for $p$ odd 
\begin{equation}\label{param_R_odd}
\nu_{odd} = {7-p\over 2}\,, \;\;\;\; \alpha_{odd}
  =  {p-1\over 2}\,, \;\;\;\;
\beta_{odd}  = 0\,.
\end{equation}
The dimensionless parameters $\wn$, $\qn$ are defined as
\begin{equation} \label{wnqn}
\wn \equiv {\nu \, \omega\over 4 \, \pi\,  T}  \,, \hspace{2cm} 
\qn \equiv {\nu\,  q \over 4 \, \pi \, T} \,.
\end{equation}
Near the horizon, two local independent solutions of Eq.~({\ref{com_odd_1}) 
are $A_t' \sim (u-1)^{a_{\pm}}$, where $a_{\pm} = \pm i\wn / \nu$.
The ``incoming wave'' boundary condition corresponds to choosing 
$a_-$ as the correct exponent. Solving Eq.~({\ref{com_odd_1}) 
perturbatively in $\wn$, $\qn$ we find
\begin{equation}
A_t' = C (u-1)^{-{ i \wn\over \nu} } u^{-\beta}
 \left( 1 + \wn \, F^{(p)}(u)
+\qn^2 \, G^{(p)}(u) +\cdots\right)\,, 
\end{equation}
here  $F^{(p)}(u)$ and $G^{(p)}(u)$ are explicitly known (but rather 
cumbersome) functions, independent of $\wn$, $\qn$.
The constant $C$ can be found from Eq.~(\ref{max2}) and the boundary 
values of fields at $u=0$.  We have
\begin{equation}\label{max2pole}
C e^{{\pi  \wn \over \nu}} \left( i \, \wn + {7-p\over 2 \nu}
\, \qn^2 + \cdots \right) =  
\left( \wn \, \qn\,  A_z^0 + \qn^2 \, A_t^0 \right)\,,
\end{equation}
where
the ellipses represent terms of higher order in $\wn$, $\qn$.
Here we first see the emergence of the diffusion pole: $C\sim
(i\wn+\frac{7-p}{2\nu} \qn^2)^{-1}$.  This pole will appear in the
correlators.  It corresponds to the diffusion constant
\begin{equation}\label{diffusion_R}
  D = \frac{7-p}{8 \pi T}\,.
\end{equation}

\subsection{Shear viscosity}
\label{shear_mode}

The low-energy string theory equations of motion which give rise to
the black $p$-brane backgrounds (\ref{es_metric}) can be deduced from
the following Einstein-frame effective action
\begin{equation}\label{grav_action}
S = \int d^{10} x \sqrt{-g} \left( {\cal R} - \frac12 \partial_{\mu} 
\Phi \partial^{\mu} \Phi - \frac1{2 n!} e^{a \Phi} F_n^2 \right)\,.
\end{equation}
Here $\Phi$ is the dilaton, $F_n$ is the Ramond-Ramond field strength
form (or its dual), and values of $n$ and $a$ are related to $p$ for 
a given $p$-brane background. The
equations of motion that follow from Eq.~(\ref{grav_action}) are 
\begin{subequations}\label{fluc}
\begin{eqnarray}
& & {\cal R}_{\mu\nu} = \frac12\, \partial_{\mu} \Phi\, \partial_{\nu} \Phi
+\frac1{2 (n-1)!}\, e^{a \Phi}\, 
\left( F_{\mu \dots} F_{\nu}^{\dots} - \frac{n-1}{8 n} F^2 \, 
g_{\mu\nu}\right)\,,\label{fluc1} \\
& & \nabla_\mu  \left( \, e^{a \Phi}\,  F^{\mu \dots}\right) = 0\,,
\label{fluc2} \\
& & \Box \Phi = \frac a{2 n!} \, e^{a \Phi}  F^2\,.\label{fluc3}
\end{eqnarray}
\end{subequations}

To find the viscosity, we can substitute the perturbed metric
$g_{\mu\nu} + h_{\mu\nu}$ 
 into Eq.~(\ref{fluc1}), solve the resulting equations,
and then use the Minkowski AdS/CFT correspondence to find the retarded
correlators of the appropriate components of the stress-energy tensor
in the dual gauge theory. The case of $p=3$ was treated in this way in
Ref.~\cite{PSShydro}.
 
Alternatively, we may use the fact that, as noted in section
\ref{shear_mp}, the shear gravitational perturbations satisfy Maxwell
equations with a coordinate-dependent effective coupling in a
dimensionally reduced background.  Thus the problem is reduced to the
one treated in Section \ref{R_charge}: the system of the effective
Maxwell's equations is given by (\ref{max1})--(\ref{max3}), with the
metrics   (\ref{es_metric_u_even}), (\ref{es_metric_u_odd}) 
for even and odd $p$ respectively, 
 $\sqrt{-g}$ replaced by $\sqrt{-g}/\gYM^2 = g_{xx}\sqrt{-g}$
and $A_t= (g_{xx})^{-1} h_{ty}$,  $A_x = (g_{xx})^{-1} h_{xy}$.  
Then the equation for the effective $A_t'$ 
 still has the form (\ref{com_odd_1}), but with a new set of
parameters,
\begin{equation}\label{param_grav_even}
 \nu_{even} = 7 - p\,, \;\;\;\; \alpha_{even}
  =  3\,, \;\;\;\;
\beta_{even}  = p-6\,,
\end{equation}
\begin{equation}\label{param_grav_odd}
\nu_{odd} = {7-p\over 2}\,, \;\;\;\; \alpha_{odd}
  =  2\,, \;\;\;\;
\beta_{odd}  = {p-5\over 2}\,.
\end{equation}
The subsequent calculations follow the Maxwell case verbatim, and in
the end we find the hydrodynamic pole in shear part of the correlator
of $T^{\mu\nu}$ at $\wn = - i \qn^2 /\nu + \cdots$.  The dispersion
relation is $\omega = - i{\cal D} q^2+\cdots$, where ${\cal D} = 1/4\pi T$
is independent of $p$, hence $\eta/s = 1/4\pi$.


\section{Discussion}
\label{sec:concl}

Let us briefly summarize.
The analysis of small fluctuations around a
black-brane background shows that in the low frequency, long-wavelength limit, 
 there
exist perturbations that correspond to diffusive hydrodynamic
processes on the stretched horizon.  This result indicates that
the degrees of freedom holographically dual to a gravitational theory on a 
black-brane background (whatever their microscopic
nature is) must have a rather conventional hydrodynamic limit at large
distances and long times.

We have also derived general expressions for the diffusion constant and shear
viscosity in terms of the components of the background metric.  For 
near-extremal
D$p$ and M-branes, these formulas reproduce the results of the direct AdS/CFT
calculations.
\footnote{
     It should be stressed that the notion of 
     diffusion (more generally, hydrodynamics) 
     on stretched horizons is not limited to 
     the particular string-theoretic realization 
     of holography.
     It is encoded in the low-frequency part of
     the spectrum of quasinormal modes for a 
     given gravitational background (see \cite{Nunez:2003eq} 
     for a discussion of quasinormal modes in this context).}
 For all non-extremal supergravity backgrounds
considered in the paper we found that the ratio of shear viscosity to
entropy density is a universal number%
\footnote{
     This observation traces back to the observation
     in \cite{Herzog} that shear viscosities and
     entropies of M-branes have the same $N$ 
     dependence.
     Curiously, the viscosity to entropy ratio is also equal
      to $1/4\pi$ in the pre-holographic ``membrane
     paradigm'' hydrodynamics \cite{Thorne:iy,Parikh:1997ma}:
     there, for a four-dimensional Schwarzschild black hole one has 
     $\eta_{\,\mathrm{m.p.}} = 1/16\pi G_N$, while 
     the Bekenstein-Hawking entropy is $s = 1/4G_N$.}
equal to $(4\pi)^{-1}$.

We find these results rather intriguing,
and some future work is yet to be done.
First,
at the moment we lack crisp understanding of why
our analysis of small perturbations in the
vicinity of the horizon yields the same result
as the AdS/CFT calculation.

Another question to be addressed is the complete
treatment of gravitational perturbations.
Indeed, the fluctuations of the stretched
horizon must encode the full set of hydrodynamic
modes, at least in the known AdS/CFT examples.
This means that methods similar to those used
in Section~\ref{secone} should reproduce
propagating modes (sound), in addition to
the non-propagating shear and diffusive modes.%
\footnote{
     Sound waves from the AdS/CFT perspective were 
     considered in \cite{PSShydro2,Herzog:2003ke}.}
In particular, it is reasonable to expect 
that formulas analogous to (\ref{eta})
exist also for bulk viscosity and for the 
speed of sound.

It would be interesting to extend and generalize our calculations to
include other nontrivial supergravity backgrounds, such as rotating
branes~\cite{Gubser:1998jb}, non-extremal Klebanov-Tseytlin
solution~\cite{Buchel:2001gw}, or the full (as opposed to the
high-temperature limit) Buchel-Liu metric \cite{Buchel:2003ah}.

Another interesting question is the gravitational
description of non-liner terms in hydrodynamic 
equations.
Namely, the hydrodynamic
constitutive relations for conserved currents
contain in addition to the linear term (Fick's law),
a non-linear convective term:
$j^i = -D \, \partial^i j^0 + (\epsilon + P)^{-1} \, j^0\, \pi^i$.
Here $\pi^i$ is momentum density in the $i$-th 
direction; the last term is just charge density 
times velocity of the fluid.
This term, which is quadratic in small fluctuations,
is a direct consequence of Lorentz (or Galilean)
invariance of the microscopic theory.%
\footnote{
     Analogous non-linear terms also exist in the
     constitutive relations for stress $T^{ij}$.}
The gravitational manifestation of the convective
term may not be straightforward: it was shown 
in \cite{KY} that such non-linear terms give 
rise to $O(1/N^2)$ effects in correlation 
functions of conserved currents.
In the context of AdS/CFT, this implies that
gravity loop effects must be included.

Finally, we hope that analyzing the hydrodynamic regime of
black branes will lead to a better conceptual understanding of
holography.

\appendix
\section{Dimensional reduction}
\label{app:dimensional reduction}
In this Appendix we write down the
formulas of the dimensional reduction scheme relevant 
for our discussion in the main text. More details can be found in 
\cite{Cho:1975sf,Scherk:1979zr,KK book,Bremer:1998zp}.

The dimensional reduction of a $D$-dimensional pure 
Einstein theory is performed by using the ansatz
\begin{equation}
ds^2 = e^{2 a\phi} d s^2_X  +  e^{2 b\phi} d s^2_Y\,,
\label{metric_D}
\end{equation}
where $X$ is the lower-dimensional space 
and $Y$ is the internal compactifying space with dimensions 
 $d_X$ and $d_Y$ respectively ($d_X + d_Y = D$).
In the reduced action, the lower-dimensional Einstein-Hilbert term and 
the kinetic term for the scalar $\phi$ will both 
have canonical normalization, if
we take 
\begin{equation}
a = \sqrt{ { d_Y \over 2 (d_X-2) (d_X+d_Y-2)}}\,,  \;\;\;\;\;\;
b = -  \sqrt{ {d_X-2\over 2 d_Y (d_X+d_Y-2) }}\,.
\end{equation}
Considering vector-like fluctuations of the metric (\ref{metric_D}) 
(gravitons with one index along the $X$ space and one index along the $Y$ one),
one observes that the normalization of the corresponding gauge fields 
is {\em not} canonical.
Schematically, 
\begin{equation}
\int \sqrt{-g}\,  {\cal R} \; \rightarrow \;
\int \sqrt{-g_X} \, e^{2(b-a)\phi } \, F_{\mu\nu}^2  \, + \cdots \,,
\end{equation}
where constant prefactors are ignored. Thus in general the 
effective gauge coupling $1/\gYM^2 \sim  e^{2(b-a)\phi }$ will be 
position-dependent.

\section{Hydrodynamics}
\label{app:hydro review}

To make our presentation self-contained,
here we briefly review
the basic properties of hydrodynamic
fluctuations. Hydrodynamics is an 
effective theory, which describes the dynamics of
a thermal system on length and time scales which are large compared to
any relevant microscopic scale.  The degrees of freedom entering this
theory, in the simplest cases, are the densities of conserved charges.

As an example, consider a translationally
invariant theory in flat space, which has a 
conserved current $j^\mu$, which corresponds 
to some global symmetry.
The hydrodynamic degrees of freedom are
the charge densities $j^0$, $\varepsilon\equiv T^{00}$,
and $\pi^i\equiv T^{i0}$, where $T^{\mu\nu}$
is the energy-momentum tensor.
The currents satisfy the conservation laws:
\begin{eqnarray}
     && \d_t j^0 = - \d_i j^i \ ,\\
     && \d_t \varepsilon = -\d_i \pi^i \ ,\\
     && \d_t \pi^i = -\d_j T^{ij} \ .
\end{eqnarray}
The constitutive relations express the spatial 
currents $j^i$, $T^{ij}$ in terms of
$j^0$, $\varepsilon$, and $\pi^i$.  To
 linear order
\begin{eqnarray}
     && j_i = - D \, \d_i j^0 \ ,\label{eq:Ficks law}\\
     && T_{ij} = \delta_{ij} \left( \P + v_s^2 \delta\varepsilon \right) 
                 -\frac{\zeta}{\epsilon+\P} \delta_{ij} \, \d_k\pi_k
                 -\frac{\eta}{\epsilon+\P} \left( 
                 \d_i\pi_j + \d_j\pi_i 
                 - \frac 2p \delta_{ij}\, \d_k\pi_k\right)\,.
\end{eqnarray}
Here $\epsilon = \langle\varepsilon\rangle$,
$\P = \frac 1p \langle T_{ii} \rangle$ are
equilibrium energy density and pressure,
$p$ is the number of spatial dimensions,
$v_s = (\d \P/\d\epsilon)^{1/2}$ is the speed
of sound, and
$\delta\varepsilon = (\varepsilon - \epsilon)$
is the fluctuation in energy density.
The unknown kinetic coefficients $D$, $\zeta$
and $\eta$ are diffusion constant, 
bulk and shear viscosities, respectively.
The constitutive relation (\ref{eq:Ficks law})
is often called Fick's law.

The linearized constitutive relations together with 
conservation laws give the dispersion 
relations for hydrodynamic modes.
Taking all charge densities to be proportional
to $e^{-i\omega t + i {\bm q}\cdot {\bm x}}$, 
one finds:
{\em i)} diffusive mode, $\omega = -i{\bm q}^2 D$,
{\em ii)} shear mode, $\omega = -i {\bm q}^2 \eta/(\epsilon+P)$, and
{\em iii)} sound mode, 
$\omega = v_s q -i {\bm q}^2 (\zeta+\frac{2p-2}{p}\,\eta)/(\epsilon+P)$.

Finally, linear response theory allows one to 
extract kinetic coefficients 
from singularities in equilibrium correlation
functions of the corresponding conserved currents.
This was used to find both diffusion 
constant and shear viscosity in some 
strongly-coupled theories from AdS/CFT 
recipe for correlation functions
\cite{PSShydro,Herzog}.
We use this method in section~\ref{ads}.

\section{Shear mode damping constant for the Buchel-Liu metric}
\label{app:Buchel-Liu}

Non-extremal deformation of the Pilch-Warner RG flow recently 
found by Buchel and Liu \cite{Buchel:2003ah} is a gravity dual 
to the finite-temperature 
${\cal N}=2^*$ $SU(N)$
gauge theory at large $N$ and large 't Hooft coupling.
The five-dimensional metric is of the form
\begin{equation}
ds^2 = e^{2A} \left( - e^{2 B} dt^2 + d\x^2\right) + dr^2\,, 
\end{equation}
where $r$ is the radial coordinate in five dimensions. Functions 
$A(r)$, $B(r)$ satisfy supergravity equations of motion governing the flow. 
The system of equations (which also involves two $r$-dependent scalars) 
is given explicitly by Eqs.~(3.17) of \cite{Buchel:2003ah}.
Following \cite{Buchel:2003ah} we introduce a new radial variable, 
$y=e^B$, $y\in [0,1]$, with $y=0$ corresponding
to the position of the horizon.
The metric becomes
\begin{equation}
ds^2 = e^{2A} \left( - y^2 dt^2 + d\x^2\right) + 
dr^2 \left( {\partial r \over \partial y}\right)^2\,. 
\label{buchel_y}
\end{equation}
Using  Eqs.~(3.17) of \cite{Buchel:2003ah} one finds the following
expression involving the Jacobian $\partial y /\partial r$
\begin{equation}
 \left( {\partial y \over \partial r}\right)^2 \left(  (A')^2 +
{1\over 2 y} A' 
- \left( {\rho '\over \rho}\right)^2-{1\over 3} (\chi ')^2\right)
= - {1\over 3} {\cal P}\,,
\end{equation}
where  prime denotes the derivative with respect to $y$. Functions
$\rho (y)$ and $\chi (y)$ are the two scalars satisfying 
supergravity equations of motion, and ${\cal P}$ is their potential,
\begin{equation}
{\cal P } = {1\over 12} \left( {1\over \rho^2} - \rho^4 \cosh{2\chi} \right)^2
+{1\over 16} \rho^8 \sinh^2 2\chi - 
 {1\over 3} \left( {1\over \rho^2} + {\rho^4\over 2}  \cosh{2\chi} \right)^2\,.
\label{jak}
\end{equation}
Applied to metric (\ref{buchel_y}), our formula (\ref{smdcd}) 
for the shear mode
damping constant reads
\begin{equation}
{\cal D} = e^{3 A(0)} \int\limits_{0}^{1} { y e^{-4 A(y)}
 \over | \partial y / \partial r|}\,. 
\label{buchel_D}
\end{equation}
Knowing $A(y)$, $\rho (y)$ and $\chi (y)$, one can in principle determine 
${\cal D}$ using Eqs.~(\ref{jak}) and (\ref{buchel_D}). 
Unfortunately, the system of 
second order differential equations involving those functions is too 
complicated. Buchel and Liu  \cite{Buchel:2003ah} 
were able to solve it perturbatively
using $\alpha_1\propto (m_b/T)^2\ll 1$ and $\alpha_2\propto m_f/T \ll 1$, 
where $m_b$ and $m_f$ are masses of the bosonic and fermionic 
components of the ${\cal N}=2$ hypermultiplet, as small parameters.
To the leading order in $\alpha_1$, $\alpha_2$ the solution 
is  \cite{Buchel:2003ah}
\begin{equation}
\begin{split}
A(y)&={\hat \alpha}-\frac{1}{4} \ln (1-y^2)+\alpha_1^2 
A_1(y)+\alpha_2^2 A_2(y)\,,\\
\rho(y)&=1+\alpha_1 \rho_1(y)\,,\\
\chi(y)&=\alpha_2 \chi_2(y)\,,
\end{split}
\label{anz}
\end{equation}
where the scalars $\rho_1 (y)$ and $\chi_2 (y)$
obey linear differential equations
\begin{eqnarray}
&\,& (1-y^2)^2\,\left(y\,\rho_1'\right)'+y\,\rho_1 =0\,,\label{eo11}\\
&\,& (1-y^2)^2\,\left(y\,\chi_2'\right)'+\frac34\,y\,\chi_2 =0\,,
\label{eo12}
\end{eqnarray}
and functions 
 $A_1(y)$, $A_2(y)$ can be found by solving 
\begin{eqnarray}
&\,& 
y (1-y^2)\,A_1''-(1+3 y^2)\,A_1'+4 y (1-y^2)\,\left(\rho_1'\right)^2=0\,,
\label{eo21}\\
&\,& y (1-y^2)\,A_2''-(1+3 y^2)\,A_2'+\frac 43 y (1-y^2)
\,\left(\chi_2'\right)^2=0\,.
\label{eo22}
\end{eqnarray}
The scalars  $\rho_1 (y)$ and $\chi_2 (y)$ are expressed in 
terms of the hypergeometric function
\begin{eqnarray}
&\,& \rho_1=(1-y^2)^{1/2}\ _2F_1({1\over2},{1\over 2},1; y^2)\,,\label{ansr2_a}\\
&\,& \chi_2=(1-y^2)^{3/4}\ _2F_1({3\over 4},{3\over 4},1;y^2)\,,
\label{ansr2_b}
\end{eqnarray}
and functions $A_{1}$, $A_2$ are given by
\begin{equation}
\begin{split}
A_1&=\xi_1-4\int_0^y\,\frac{z\,dz}{(1-z^2)^2}\
\biggl( {8-\pi^2\over 2 \pi^2}
+\int_0^z dx\,\left(\frac{\partial\rho_1}{\partial x}
\right)^2\,
\frac{(1-x^2)^2 }{x}\biggr)\,,\\[4pt]
A_2&=\xi_2-\frac 43 \int_0^y\,\frac{z\,dz}{(1-z^2)^2}\
\biggl( {8-3\pi\over 8 \pi}
+\int_0^z dx\,\left(\frac{\partial\chi_2}{\partial x}
\right)^2\,
\frac{(1-x^2)^2 }{x}\biggr)\,.\\
\end{split}
\label{ansr34}
\end{equation}
To the leading order in $\alpha_1$, $\alpha_2$, 
integration constants ${\hat \alpha}$, $\xi_1$, $\xi_2$ are related 
to the Hawking temperature of the black brane background by
\begin{equation}
T = {1\over 2\pi} e^{{\hat \alpha}} 
\left[ 1 + \alpha_1^2 \left( \xi_1 + {16\over \pi^2}\right) 
+\alpha_2^2 \left( \xi_2 +{4\over 3\pi}\right)\right]\,.
\label{TTf}
\end{equation}
For  $\alpha_1=0$, $\alpha_2=0$ one recovers the original non-extremal 
black three-brane metric
\begin{equation}
ds^2 = (2\pi T)^2(1-y^2)^{-1/2} \left( - y^2 dt^2 + d\x^2\right) +
{dy^2\over (1-y^2)^2}\,.
\label{ansr34a}
\end{equation}
Expanding the Jacobian to the leading order in  $\alpha_1$, $\alpha_2$ we find
\begin{eqnarray}
 {\partial y \over \partial r} &=& (1-y^2) \Biggl[ 1 +\alpha_1^2
\left( 2 \rho_1^2 + 2 (1-y^2)^2 (\rho_1')^2 - {1-y^4\over y } A_1' \right)
\nonumber \\ 
&+& 
\alpha_2^2 \left( {1\over 2} \chi_2^2 + {2\over 3} (1-y^2)^2 (\chi_2')^2 - 
 {1-y^4\over y } A_2'\right) \Biggr]\,.
\end{eqnarray}
Correspondingly, the shear mode damping constant (\ref{buchel_D}) 
can be written as a series expansion 
\begin{equation}
{\cal D} = {1\over 2 \pi T}
 \int\limits_{0}^{1} dy y \left( 1 + 
\alpha_1^2 F_1(y) + \alpha_2^2 F_2(y) +\cdots\right)\,,
\end{equation}
where ellipses stand for terms of higher order 
in $\alpha_1$, $\alpha_2$, and
 \begin{eqnarray}
 F_1 &=& {1-y^4\over y } A_1' - 4 {\bar A_1} 
- 2 \rho_1^2 - 2 (1-y^2)^2 (\rho_1')^2 +{16\over \pi^2}\,, \label{FF1}\\
F_2 &=& {1-y^4\over y } A_2' - 4 {\bar A_2} -
 {1\over 2} \chi_2^2 - {2\over 3} (1-y^2)^2 (\chi_2')^2 +{4\over 3\pi}\,,
\label{FF2}
\end{eqnarray}
where  ${\bar A_1} = A_1 (\xi_1=0)$,  ${\bar A_1} = A_2 (\xi_2=0)$.
Using equations of motion (\ref{eo11}) - (\ref{eo12}), (\ref{eo21})-
(\ref{eo22}) we now show that
$F_1$ and $F_2$ vanish identically.

To prove that $F_1\equiv 0$, first we integrate Eq.~(\ref{eo21})
from $0$ to $y$ (taking into account that ${\bar A_1}(0)=0$, 
$A_1'(0)=0$) to obtain the identity
\begin{equation}
{\bar A_1} (y) = {1\over 2} y (1-y^2) A_1' + 
 2 \int\limits_{0}^{y} z(1-z^2) (\rho_1')^2 dz\,.
\end{equation}
Inserting this into Eq.~({\ref{FF1}) we find 
\begin{equation}
F_1 = 1 - \rho_1^2 - (1-y^2)^2 (\rho_1')^2 - 2
 \int\limits_{0}^{y} {1-z^4\over z} (\rho_1')^2 dz\,.
\label{idrho}
\end{equation}
To see that the right hand side of Eq.~(\ref{idrho}) is identically zero, 
we multiply both sides of the differential equation  (\ref{eo11}) by 
$\rho_1'$. We have then
\begin{equation}
(\rho_1')^2 = - {y (\rho_1^2)'\over (1-y^2)^2} - {y \over 2} [(\rho_1')^2]'\,.
\label{addequal}
\end{equation}
This allows us to express the integrand in Eq.~({\ref{idrho}) as
\begin{eqnarray}
{2 (1-z^4)\over z} (\rho_1')^2 & = & - {1+z^2\over 1-z^2} (\rho_1^2)' -
(1-z^4)[(\rho_1')^2]' \nonumber \\
&=& -  (\rho_1^2)' -
(1-z^2)^2 [(\rho_1')^2]'  + 4 z (1-z^2)(\rho_1')^2 \nonumber \\
&-&    {2 z^2\over 1-z^2} (\rho_1^2)' -
2 z^2(1-z^2) [(\rho_1')^2]'  - 4 z (1-z^2)(\rho_1')^2\,.
\label{longeq}
\end{eqnarray}
 Eq.~(\ref{addequal}) ensures that the
 last line in Eq.~({\ref{longeq}) adds up to zero.
Noting that 
\begin{equation}
 -  (\rho_1^2)' -
(1-z^2)^2 [(\rho_1')^2]'  + 4 z (1-z^2)(\rho_1')^2 
= {d \over d z}\left( - \rho_1^2 - (1-z^2)^2 (\rho_1')^2 \right) \,
\label{addequal1}
\end{equation}
and remembering that $\rho_1(0)=1$, $\rho_1'(0)=0$ (see (\ref{ansr2_a})), 
we arrive at 
\begin{equation}
 2 \int\limits_{0}^{y} {1-z^4\over z} (\rho_1')^2 dz = 
 1 - \rho_1^2 - (1-y^2)^2 (\rho_1')^2\,.
\label{addequal2}
\end{equation}
Consequently, $F_1\equiv 0$. The proof of  $F_2\equiv 0$ is similar.
We conclude therefore 
that (at least) to the next to 
leading order in the high temperature expansion 
 parameters $m_b/T$, $m_f/T$
the shear mode damping constant of the ${\cal N}=2^*$ 
gauge theory at large $N$ and large 't~Hooft coupling
is independent of the parameters of deformation (hypermultiplet masses), 
and equals $1/4\pi T$.

\acknowledgments 

P.K.K. thanks A.~Karch and L.~Yaffe 
for discussions and comments.
A.O.S. would like to thank D.~Birmingham, O.~DeWolfe,
 G.~Horowitz, L.~Lindblom, A.~Mikhailov, J.~Polchinski,
R.~Roiban, A.~Volovich, J.~Walcher, and especially A.~Buchel and C.~Herzog 
 for conversations, questions and comments,
and the KITP at UC Santa-Barbara for its warm hospitality.  P.K.K.\ is
supported, in part, by DOE Grant No.\ DE-FG03-96ER40956.  D.T.S.\ and
A.O.S.\ are supported, in part, by DOE Grant No.\ DOE-ER-41132.  The work of
D.T.S.\ is supported, in part, by the Alfred P.\ Sloan Foundation.


\begin{thebibliography}{99}

\bibitem{'tHooft:gx}
G.~'t Hooft,
{\em ``Dimensional reduction in quantum gravity,''}
[\grqc{9310026}].

\bibitem{Susskind:1994vu}
L.~Susskind,
{\em ``The world as a hologram,''}
\jmp{36}{1995}6377
[\hepth{9409089}].

\bibitem{Bousso:2002ju}
R.~Bousso,
{\em ``The holographic principle,''}
\rmp{74}{2002}{825}
[\hepth/0203101].

\bibitem{Maldacena:1997re}
J.~Maldacena,
{\em ``The large N limit of superconformal field theories and supergravity,''}
\atmp{2}{1998}{231}
[\hepth{9711200}].

\bibitem{Gubser:1998bc}
S.S.~Gubser, I.R.~Klebanov and A.M.~Polyakov,
{\em ``Gauge theory correlators from non-critical string theory,''}
\plb{428}{1998}{105}
[\hepth{9802109}].

\bibitem{Witten:1998qj}
E.~Witten,
{\em ``Anti de Sitter space and holography,''}
\atmp{2}{1998}{253}
[\hepth{9802150}].

\bibitem{Aharony:1999ti}
O.~Aharony, S.S.~Gubser, J.~Maldacena, H.~Ooguri and Y.~Oz,
{\em ``Large N field theories, string theory and gravity,''}
\prep{323}{2000}{183} 
[\hepth{9905111}].

\bibitem{Thorne:iy}
K.S.~Thorne, R.H.~Price and D.A.~Macdonald,
{\em ``Black holes: the membrane paradigm,''}
Yale University Press, New Haven 1986.

\bibitem{Parikh:1997ma}
M.~Parikh and F.~Wilczek,
{\em ``An action for black hole membranes,''}
\prd{58}{1998}{064011}
[\grqc{9712077}].

\bibitem{Policastro:2001yc}
G.~Policastro, D.T.~Son and A.O.~Starinets, 
{\em ``Shear viscosity of strongly coupled ${\cal N} = 4$ supersymmetric 
Yang-Mills plasma,''}
\prl{87}{2001}{081601}
[\hepth{0104066}].

\bibitem{PSShydro}
G.~Policastro, D.T.~Son and A.O.~Starinets,
{\em ``From AdS/CFT correspondence to hydrodynamics,''}
\jhep{09}{2002}{043}
[\hepth{0205052}].

\bibitem{Herzog}
C.P.~Herzog,
{\em ``The hydrodynamics of M-theory,''}
\jhep{12}{2002}{036}
[\hepth{0210126}].

\bibitem{PSShydro2}
G.~Policastro, D.T.~Son and A.O.~Starinets,
{\em ``From AdS/CFT correspondence to hydrodynamics, II: Sound waves,''}
\jhep{12}{2002}{054}
[\hepth{0210220}].

\bibitem{Herzog:2003ke}
C.~P.~Herzog,
``The sound of M-theory,''
\prd{68}{2003}{024013}
[arXiv:hep-th/0302086].



\bibitem{KY}
P.~Kovtun and L.~G.~Yaffe,
{\it ``Hydrodynamic fluctuations, long-time tails, and supersymmetry,''}
\prd{68}{2003}{025007}
[\hepth{0303010}].

\bibitem{Son:2002sd}
D.T.~Son and A.O.~Starinets,
{\em ``Minkowski space correlators in AdS/CFT correspondence: recipe 
and applications,''}
\jhep{09}{2002}{42}
[\hepth{0205051}].


\bibitem{Herzog:2002pc}
C.~P.~Herzog and D.~T.~Son,
``Schwinger-Keldysh propagators from AdS/CFT correspondence,''
\jhep{0303}{2003}{46}
[arXiv:hep-th/0212072].



\bibitem{Horowitz:cd}
G.T.~Horowitz and A.~Strominger,
{\em ``Black strings and p-branes,''}
\npb{360}{1991}{197}.

\bibitem{Buchel:2003ah}
A.~Buchel and J.T.~Liu,
{\em ``Thermodynamics of the ${\cal N} = 2^*$ flow,''}
[\hepth{0305064}].

\bibitem{Pilch:2000ue}
K.~Pilch and N.~P.~Warner,
{\em ``$N = 2$ supersymmetric RG flows and the IIB dilaton,''}
\npb{594}{2001}{209}
[\hepth{0004063}].

\bibitem{CRC}
CRC handbook of chemistry and physics, 83rd edition,
CRC Press, 2002-2003.

\bibitem{Gherghetta:2001iv}
T.~Gherghetta and Y.~Oz,
{\em ``Supergravity, non-conformal field theories and brane-worlds,''}
\prd{65}{2002}{046001}
[\hepth{0106255}].

\bibitem{Klebanov:1996un}
I.R.~Klebanov and A.A.~Tseytlin,
{\em Entropy of near-extremal black p-branes},
\npb{475}{1996}{164}
[\hepth{9604089}].

\bibitem{Chalmers:1998xr}
G.~Chalmers, H.~Nastase, K.~Schalm and R.~Siebelink,
{\em R-current correlators in $N = 4$ super Yang-Mills theory 
from anti-de  Sitter supergravity},
\npb{540}{1999}{247}
[\hepth{9805105}].

\bibitem{Nunez:2003eq}
A.~Nunez and A.~O.~Starinets,
``AdS/CFT correspondence, quasinormal modes, and thermal correlators in  N = 4 SYM,''
 \prd{67}{2003}{124013}
[arXiv:hep-th/0302026].


\bibitem{Gubser:1998jb}
S.S.~Gubser,
{\em Thermodynamics of spinning D3-branes},
\npb{551}{1999}{667}
[\hepth{9810225}].

\bibitem{Buchel:2001gw}
A.~Buchel, C.P.~Herzog, I.R.~Klebanov, L.A.~Pando Zayas and A.A.~Tseytlin,
{\em Non-extremal gravity duals for fractional D3-branes on the conifold},
\jhep{04}{2001}{033}
[\hepth{0102105}].

\bibitem{Cho:1975sf}
Y.M.~Cho and P.G.~Freund,
{\em ``Nonabelian gauge fields as Nambu-Goldstone fields,''}
\prd{12}{1975}{1711}.

\bibitem{Scherk:1979zr}
J.~Scherk and J.H.~Schwarz,
{\em ``How to get masses from extra dimensions,''}
\npb{153}{1979}{61}.

\bibitem{KK book}
T.~Appelquist, A.~Chodos and P.~G.~Freund (eds.),
{\em ``Modern Kaluza-Klein Theories,''}
Addison-Wesley, 1987.

\bibitem{Bremer:1998zp}
M.S.~Bremer, M.J.~Duff, H.~Lu, C.N.~Pope and K.S.~Stelle,
{\em ``Instanton cosmology and domain walls from M-theory and string theory,''}
\npb{543}{1999}{321}
[\hepth{9807051}].

%








\end{thebibliography}
\end{document}